# Composition, Size, and Surface Functionalization dependent Optical Properties of Lead Bromide Perovskite Nanocrystals

Palvasha Ijaz,[1,2] Muhammad Imran,[1] Márcio M. Soares,[3] Hélio C. N. Tolentino,[3] Beatriz Martín-García,[1,4] Cinzia Giannini,[5] Iwan Moreels,[1,6] Liberato Manna*,[1] Roman Krahne*[1]

[1]Department of Nanochemistry, [4]Graphene Labs, Istituto Italiano di Tecnologia, Via Morego 30, 16163 Genova, Italy

[2]Dipartimento di Chimica e Chimica Industriale, Università degli Studi di Genova, Via Dodecaneso 31, 16146 Genova, Italy

[3]Brazilian Synchrotron Light Laboratory (LNLS), Brazilian Center for Research in Energy and Materials (CNPEM), Campinas, SP, 13083-970, Brazil

[5]Istituto di Cristallografia - Consiglio Nazionale delle Ricerche (IC-CNR), via Amendola 122/O, I-70126 Bari, Italy

[6]Department of Chemistry, Ghent University, Krijgslaan 281-S3, 9000 Gent, Belgium

Corresponding author email: roman.krahne@iit.it

ABSTRACT.

The photoluminescence (PL), color purity, and stability of lead halide perovskite nanocrystals depend critically on the surface passivation. We present a study on the temperature dependent PL and PL decay dynamics of lead bromide perovskite nanocrystals characterized by different types of A cations, surface ligands, and nanocrystal sizes. Throughout, we observe a single emission peak from cryogenic to ambient temperature. The PL decay dynamics are dominated by the surface passivation, and a post-synthesis ligand exchange with a quaternary ammonium bromide (QAB) results in a more stable passivation over a larger temperature range. The PL intensity is highest from 50K-250K, which indicates that the ligand binding competes with the thermal energy at ambient temperature. Despite the favorable PL dynamics of nanocrystals passivated with QAB ligands (monoexponential PL decay over a large temperature range, increased PL intensity and stability), the surface passivation still needs improvement toward increased emission intensity in nanocrystal films.

**TOC GRAPHICS**

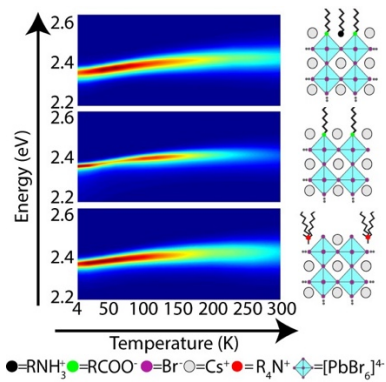

The optical properties of colloidal semiconductor nanocrystals (NCs) have been widely investigated in the last three decades. It has been well established that the composition and the nature of the NC surface strongly influence their optical properties.[1-3] Recently, lead halide perovskite NCs (LHP NCs), with $APbX_3$ composition (with A being a monovalent cation and X being either Cl, Br, or I), have emerged as a promising material due to their ease of preparation, broadly tunable band gap, high photoluminescence quantum yield (PLQY) and excellent color purity.[4-8] This remarkable set of properties makes them ideal candidates for light emission technologies, such as light emitting diodes, lasers and single photon emitters.[9-11]

Significant progress has been made on the synthesis of LHP NCs, especially with regard to size, shape, composition control, and by tailoring the surface passivation through direct synthesis or by post-synthesis ligand exchange.[4] Advancements on the synthesis of LHP NCs not only have paved the way to study shape and composition dependent optical properties, but have also offered a great opportunity to elucidate the effects of surface functionalization. PL spectroscopy at cryogenic temperatures has been used to investigate the temperature-dependent excitonic properties of traditional semiconductors and recently of LHP NCs.[12-14] In this respect, there has been some disagreement in the literature on whether LHP NCs undergo temperature-induced phase transitions from room temperature to cryogenic temperatures, on whether the emission at cryogenic temperatures consists of single or multiple peaks,[15-20] and, in the latter case, on what is the exact origin of these multiple peaks.

From a surface chemistry point of view, the first generation of LHP NCs were typically prepared by using primary alkyl amines and alkyl carboxylates as surfactants, and it was established that both ligands are present on the surface of the NCs, bound as Cs-carboxylate/alkyl ammonium bromide ligand pairs (henceforth referred to as "mixed ligands capped NCs").[21-25] Substantial advances in the colloidal synthesis and post-synthesis treatments over the past few years have provided the opportunity to prepare LHP NCs with diverse surface coatings, leading to tailored properties, such as improved colloidal stability and near-unity PLQY.[26-29] However, investigations based on optical spectroscopy were mainly limited to the first generation of NCs, *i.e.* those characterized by a mixed ligands surface passivation. Considering the recent developments in the synthesis of monodisperse NCs and in surface functionalization, a temperature dependent optical spectroscopy study of trap-free, near unity PLQY NCs should provide further insights on the effect of size, composition, and surface passivation on their excitonic properties.

In this work, we study the photoluminescence properties of $APbBr_3$ (A=Cs, MA, FA) NCs with respect to temperature, nanocrystal size, and surface passivating ligands and observe the following: (i) All samples manifest a single narrow emission peak at room and cryogenic temperatures. Since our temperature dependent XRD study excludes phase transitions, we conclude that the possible observation of multiple emission peaks at low temperatures (as was reported in the literature[16-20]) is related to polydispersity of the samples. (ii) The PL intensity is strongest in an intermediate temperature regime that spans from around 50K to 250K, while the PL lifetime decreases with decreasing temperature in the range from RT to around 50 K. PL and PL lifetimes below 50 K become strongly surface dependent and this may be ascribed to possible phase transition in the organic capping layer, as previously reported for CdSe NCs.[30-31] This behavior indicates that the ligand binding can still be improved significantly to provide the most efficient surface passivation at temperatures that are relevant for optoelectronic applications, *i.e.* at room temperature and above. (iii) The temperature induced PL red-shift[14] is more dominant in larger NCs than in smaller NCs. However, we do not observe any significant impact of NC size on the spectral shape of the emission or on the lifetime dynamics. (iv) The surface passivation affects the temperature dependence of the PL and the PL lifetime. Here, exchanging the ligands from Cs-oleate to didodecyldimethylammonium bromide (QAB) results in higher PL intensity over an extended temperature range (from 20 K to 280 K). Also, the PL decay dynamics are notably different for QAB ligands, showing a monoexponential decay over a large temperature range, while Cs-oleate and mixed ligands capped NCs develop a biexponential decay at or shortly below room temperature due to a fast non-radiative decay component. This behavior points to less effective ligand passivation at lower temperature, which we ascribe to reduced dynamics of the ligand binding at the NC surface.

Results and Discussion

The APbBr$_3$ perovskite NCs were synthesized following our benzoyl halide based procedure, as reported previously[32] (see Experimental Section in the Supporting Information (SI) for details). They were prepared using oleylamine and oleic acid as surfactants, hence they have a mixed ligand passivation (Cs-oleate and oleylammonium bromide). Transmission electron microscopy (TEM) images of CsPbBr$_3$, MAPbBr$_3$ and FAPbBr$_3$ are displayed in Figure 1 a-c. The images show that the NCs are nearly monodisperse, with roughly cubic shapes in all cases. Typical UV-Vis optical absorption and PL spectra, measured in toluene dispersions, are reported in Figure 1d, displaying a single emission peak that is red-shifted from the absorption edge.

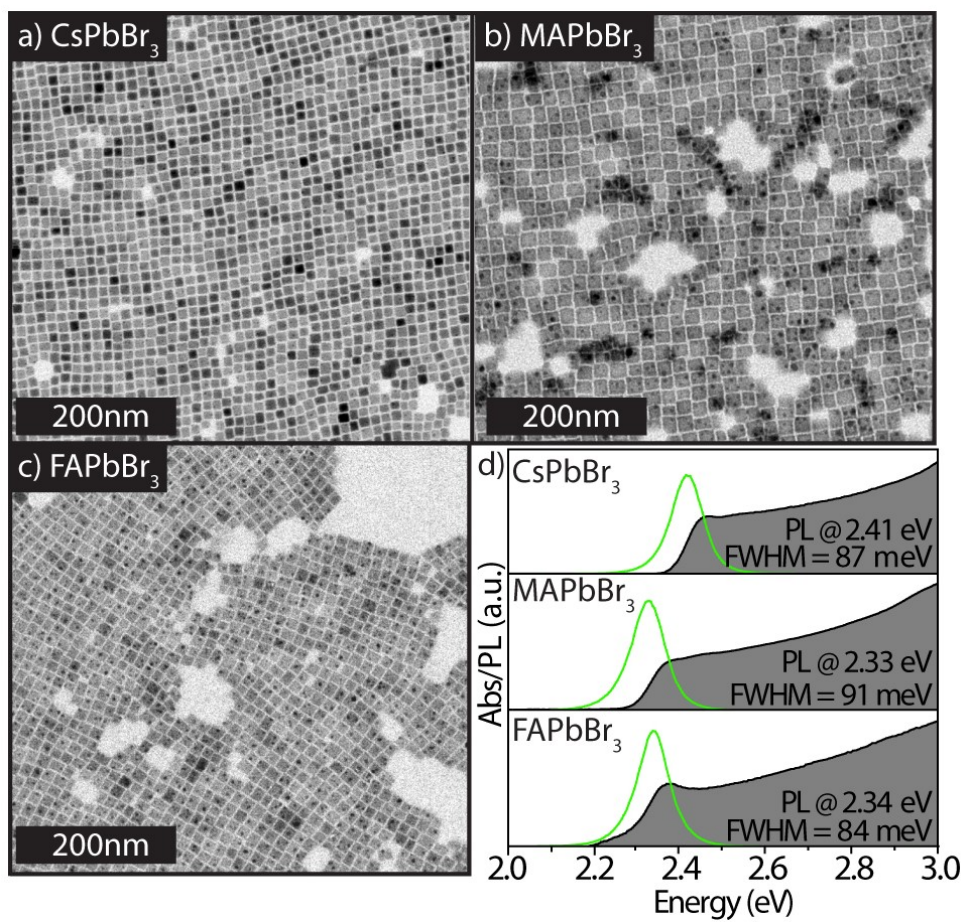

Figure 1. APbBr$_3$ NCs prepared by using oleylamine and oleic acid as surfactants, hence having a mixed ligand passivation of Cs-oleate and oleylammonium bromide (A=Cs, MA, FA). TEM images of CsPbBr$_3$ (a), MAPbBr$_3$ (b), and FAPbBr$_3$ NCs (c). The mean values of the edge lengths of the NCs in (a), (b) and (c) are 9.2 ±0.8, 17.3 ±1.6 ,12.9 ±1.4 nm, respectively. Absorbance and PL spectra of the corresponding NC samples in toluene dispersions (d).

For the temperature-dependent spectroscopic study NC films were prepared on sapphire substrates by drop-casting from the colloidal dispersions. We first focus on the samples that are characterized by different A cations and passivated by a mixed ligand shell. The films were cooled to 4 K and the temperature was increased stepwise from 4 K to 300 K to acquire the PL spectra and PL lifetime decay. The representative PL spectra recorded at 4 K and 300 K are shown in Figure 2a, whereas the complete range of PL spectra acquired at various temperatures is shown in Figures S1-S3. At T=4K, the PL peak energy is red shifted by 70, 110 and 80 meV for Cs-, MA- and FA-based perovskite NCs with respect to the corresponding spectra at RT. Such red-shift with decreasing temperature is commonly observed in lead halide perovskites, and has been attributed to a temperature dependence of the overlap between the Pb-6s and Br-4p orbitals, leading to a decrease in the band gap with decreasing temperature,[14, 33-35] and recently was found to be size dependent.[36] The MAPbBr$_3$ NCs manifest the strongest temperature related red shift in PL peak energy, which can be ascribed to their larger NC size (see also the size-dependence discussion below). The quantitative analysis showing the trends in temperature-dependent PL spectra, the PL peak energy and linewidth, and the average PL decay times are reported in Figures S4 and S5 in the SI for the three NC samples. Fitting the temperature change of the PL linewidth indicates that homogeneous broadening is dominated by coupling to LO phonons,[37] where we obtain LO phonon energies in the range from 10-40 meV (see Figure S6, Table S4 and the related discussion). The PL intensity versus temperature is plotted in Figure. 2b. For all three samples, the PL intensity is highest at intermediate temperatures (around 50-250 K) and decreases towards 4K and 300K. The PL decay traces recorded at different temperatures are reported in Figure 2c and their average PL lifetime (including fitting parameters) are reported in Tables S1-S3. At room temperature, the PL decay is almost mono-exponential, and then, with decreasing temperature, it develops a multiexponential trace with fast and slow components. For temperatures below 70K, the drop in PL intensity indicates that the non-radiative rates gain in weight.

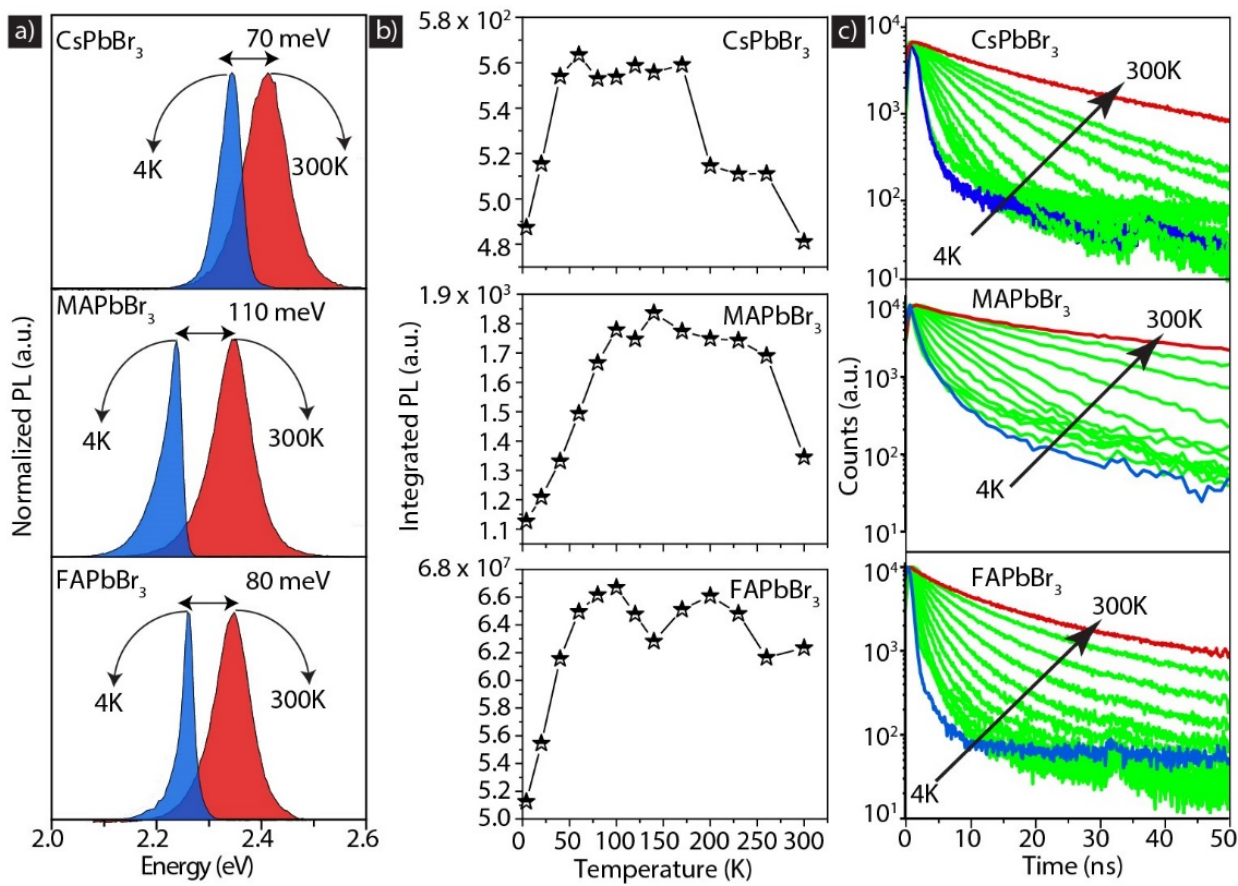

Figure 2. Photoluminescence characteristics of mixed ligands capped $CsPbBr_3$, $MAPbBr_3$ and $FAPbBr_3$ NC films. (a) Representative PL spectra recorded at 4 K and 300 K. (b, c) Integrated PL intensity (b) and PL decay traces (c) at different temperatures.

The PL emission properties of a film of perovskite NCs is expected to depend on the size dispersion of the sample, and on its surface chemistry. To investigate the impact of the surface chemistry, we decided to focus on the NC sample with Cs as A site cation ($CsPbBr_3$), and prepared two additional samples, characterized by two types of surface coatings that are different from the mixed ligand shell discussed above (Fig. 3a-c). These were Cs-oleate coated NCs in one case, and quaternary ammonium bromide (QAB) coated NCs in the other, see Figure 3. Cs-oleate coated NCs were prepared by performing the synthesis using a secondary amine instead of oleylamine.

The procedure is discussed in a previous work of ours, which demonstrated that the secondary amines are not able to bind to the surface of NCs,[38] leaving only Cs-oleate as surface coating agent. This synthesis delivers NCs with narrow size distributions and prevents the formation of low-dimensional structures (such as nanoplatelets). With this approach, we also prepared NCs with two different sizes (edge lengths of 9.5 nm and 6.4 nm) to investigate the impact of the NC size on the optical properties. QAB-coated NCs were then prepared by performing a post-synthesis ligand exchange procedure on the former samples, as also reported in a previous work from our group.[39] The ligand exchange does not affect the overall morphology and structural properties of the NCs (see TEM images of Figure S7), and there is only a slight spectral blue shift in the photoluminescence of around 10 meV (ascribed to mild surface etching[26]), while the PLQY is increased.[39] Note that Cs-oleate capped NCs are characterized by halide vacancies that are detrimental for the PLQY (which is typically below 80% for this sample), whereas both mixed ligands capped NCs and QAB capped NCs have PLQYs higher than 90%, reaching 100% for the QAB capped ones in colloidal dispersions.[32, 38-39]

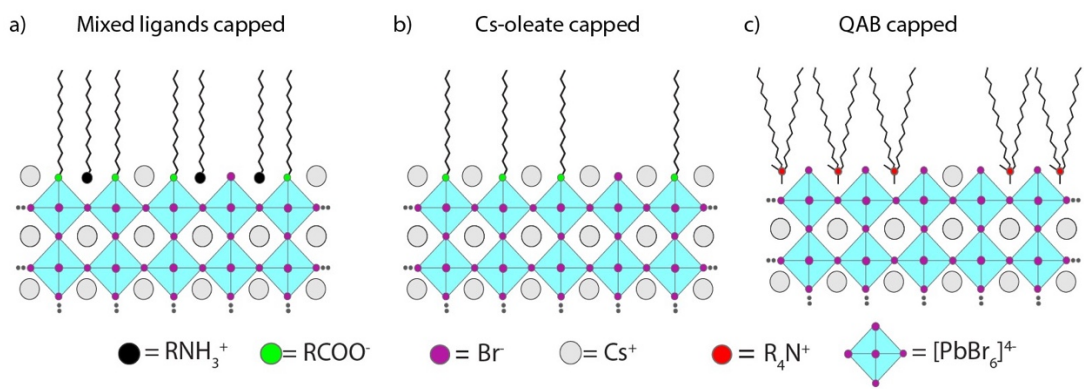

Figure 3. Schematic illustration of CsPbBr$_3$ NCs with different surface passivation: (a) mixed ligands (Cs-leate and oleylammonium bromide), (b) Cs-oleate, (c) QAB (didodecyl dimethylammonium bromide).

UV-Vis absorption and PL spectra measured from the colloidal dispersions of Cs-oleate-capped NCs with different sizes are reported in Figure 4 a,b (see Figure S7 for TEM images). The size uniformity in both samples is corroborated by the appearance of distinctive excitonic features in their optical absorption spectra (grey shaded spectra). Both samples manifest a single emission peak, with narrow PL linewidth in the range of 72-73 meV. The PL peak position depends on quantum confinement, with a blue-shifted emission for the smaller NC sample (2.45 eV at RT) with respect to the larger one (2.41 eV at RT). The PL peak for both Cs-oleate capped samples red-shifts with decreasing temperature and becomes narrower in linewidth (Figure 4 d,e,g,h). The PL amplitude is largest at low temperature, manifesting a stretched exponential tail at its low energy shoulder that can be ascribed to a broad band of defect states, probably due to halide vacancies. The red-shift with decreasing temperature is significantly reduced for the smaller nanocrystal sample, which is in agreement with the results in ref.[36]. Studies on PbS nanocrystals, that show a qualitatively similar temperature dependent band gap shift,[40] reveal the influence of the exciton binding energy, exciton-phonon coupling, and exciton fine structure at the band edge on this behavior.[41] The PL decay traces are depicted in Figure 4 j,k and show no NC size related difference. For both NC sizes, a fast decay component develops and gains in weight with decreasing temperature, while the slower component deceases in weight in the cryogenic range.

The corresponding data for the QAB coated NCs is reported in Figure 4 (c,f,i,l). Note that QAB ligands are proton-free and therefore NCs coated with these ligands are much more stable over time compared to the samples discussed above.[29, 39] Interestingly, the PL amplitude versus temperature and the PL decay traces are markedly different for the QAB passivated NCs compared to the other samples. The PL decay traces show an almost monoexponential decay that persists also at lower temperatures, down to 170 K. The PL peak of the QAB passivated sample at T=4K also has less tailing towards lower energies as compared to the other surface ligands, which indicates a lower density of trap states, thus confirming the more efficient passivation. These measurements demonstrate that the PL decay dynamics (and therefore the PL intensity) depends heavily on the surface passivation of the NC, while PL energy and line width are defined by the NC size and choice of the monovalent (A) cation. From the decay dynamics of the samples with different ligands we draw the following picture: The almost monoexponential decay at room temperature for mixed ligands and QAB passivation points to a single radiative decay channel and negligible non-radiative decay, which is corroborated by the high PL intensity. With decreasing temperature PL intensity increases and the PL decay slope becomes steeper, indicating an increase of the radiative rate of that decay channel. With further decreasing temperature a faster decay component emerges and the PL decay becomes biexponential. This can be rationalized by a slowing down of the dynamic ligand binding on the NC surface, which leads to less efficient passivation. This effect is balanced across an intermediate temperature range by the increasing rate of the radiative channel. At temperatures below 50K the non-radiative decay takes over and the PL intensity drops significantly. For Cs-oleate passivated NCs the PL decay is already biexponential at room temperature, which can be tentatively related to the presence of Br vacancies that induce non-radiative decay. This correlates well with the less efficient surface passivation of Cs-oleate leading to lower PLQY.[38]

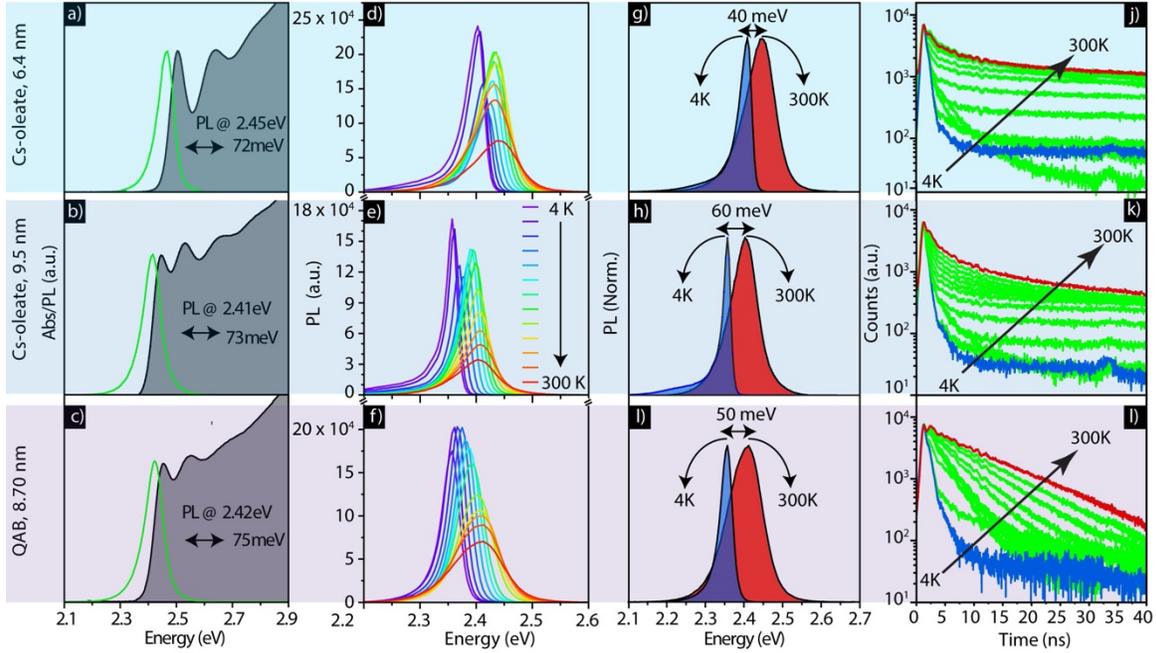

Figure 4. Optical features of CsPbBr$_3$ NCs with different sizes and surface coatings (either Cs-oleate or QAB), as indicated for each row: Absorbance and PL spectra of Cs-oleate-capped NC with different size in colloidal dispersion (a,b), and of QAB capped NCs obtained from a ligand exchange applied to the larger NC sample. (d-f) The temperature dependent PL spectra recorded on films fabricated from these samples. (g-i) Comparison of spectra recorded from films of NCs at 4 K and 300 K. Normalized PL decay traces collected for different temperatures from 4 to 300 K (j-l). PL intensities and linewidth versus temperature are reported in Figure S8.

The electronic structure, and therefore also the optical properties, are intimately related to the structure of the NC lattice, and temperature dependent phase transitions from cubic to tetragonal to orthorhombic have been reported for lead halide perovskite films.[42-46] Furthermore, it was recently demonstrated that structural defects have an impact on the phase transition in CsPbX$_3$ NCs at low temperature.[47] In order to gain insight in the structural evolution of our NCs with temperature, and in particular to test if temperature-induced phase transitions occur in our NC samples, we carried out temperature dependent X-ray diffraction (XRD) measurements on mixed ligands shell and Cs-oleate (see Figure S10 and Table S5) capped $CsPbBr_3$ NCs. Cs-oleate capped NCs inherit significant halide defects, which is further reflected in their low PLQYs in the solution phase (<80%). Figure S9 shows the experimental patterns and relative Rietveld fits in the entire temperature range for mixed ligands capped $CsPbBr_3$ NCs. The acquired data was indexed to the diffraction pattern of the $CsPbBr_3$ orthorhombic phase (ICSD code #97851), and was accordingly fitted in the entire temperature range (Figure S9). The variation with temperature of the *a*, *b* and *c* lattice parameters and unit cell volume *V*, as extracted from the Rietveld fits, are displayed in Figure 5. No phase transformation was registered in the entire temperature range, in agreement with previous reports on $CsPbX_3$ NCs.[15, 48-49] Upon cooling of the NCs films, Rietveld data analysis revealed a decrease of the *a* and *c* lattice parameters (note that *a* and *b* have almost the same value for this crystal structure, therefore our analysis attributes most of the variation to one of the two parameters, in this case a), independent of the type of surface passivation. The relevant information resides in the volume of the unit cell that decreases with temperature. These changes in the unit cell were reversible when NCs film was heated back to 300 K. Overall, apart from a smooth decrease in cell parameters and cell volume, no phase transitions were observed in both mixed ligands and Cs-oleate capped NCs.

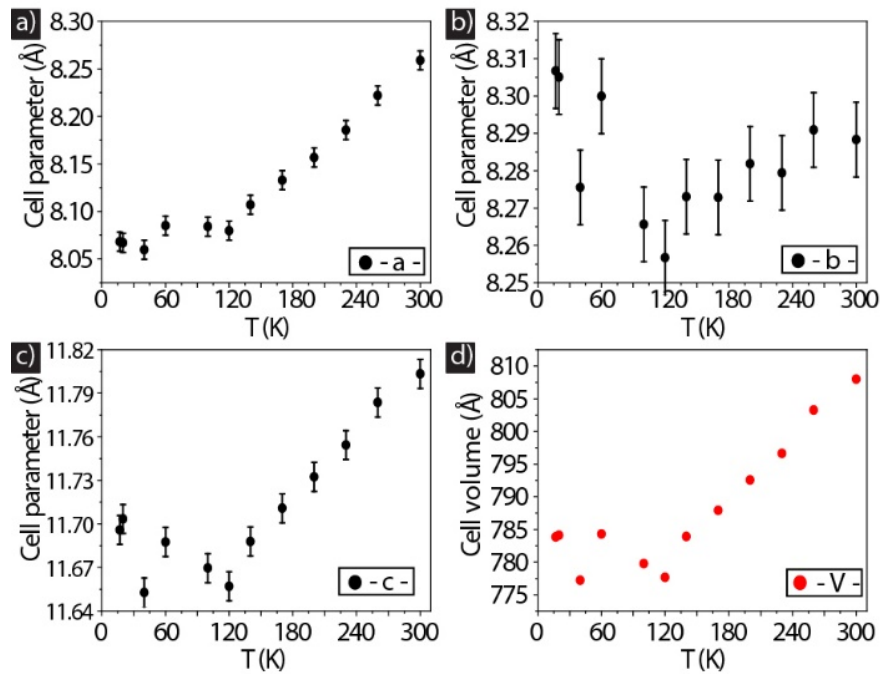

Figure 5. (a-d) Cell parameters (*a,b,c,V*) obtained from the temperature dependent XRD measurements of mixed ligands capped CsPbBr$_3$ NCs. The data yields a thermally induced expansion in the unit cell while retaining the orthorhombic phase.

In conclusion, our study on the photophysics of lead bromide perovskite nanocrystal films demonstrated that their low temperature emission is defined by a single photoluminescence peak, and that no temperature-induced phase transitions occur in these materials in the investigated temperature range. Therefore, the possible observation of multiple emission peaks at low temperature[16-20] should originate from different NC populations within the same sample. The PL peak energy, and its temperature induced shift is strongly related to the NC size, while the PL intensity and the recombination dynamics of the photoexcited carriers depend mostly on the surface functionalization. QAB ligands lead to improved PL stability and monoexponential PL decay over a larger temperature range among the three investigated types of surface passivation. The drop in PL intensity from 250K towards room temperature shows that the surface passivation of lead bromide perovskite nanocrystals needs still to be improved towards a more stable ligand binding in the range exceeding room temperature. This is particularly important for the application of such materials in optoelectronic devices.

ASSOCIATED CONTENT

Supporting Information. Synthesis protocols, PL, PL lifetime decays and their multiexponential fitting parameters, TEM images for Cs-oleate and QAB capped $CsPbBr_3$ NCs, XRD spectra recorded at different temperatures for mixed ligands and Cs–oleate capped NCs. The following files are available free of charge.

brief description (file type, i.e., PDF)

ACKNOWLEDGMENT

The research leading to these results has received funding from the European Union under the Marie Skłodowska-Curie RISE project COMPASS No.691185.